\documentclass{article}

\title{Fault-Tolerant Quantum Computation with Local Gates}

\author{Daniel Gottesman\thanks{e-mail: gottesma@t6-serv.lanl.gov}\\
T-6 Group, Los Alamos National Lab, Los Alamos, NM 87545 USA}

\date{}

\newcommand{\ket}[1]{|#1\rangle}
\newsavebox{\xline}
\newsavebox{\oline}
\newsavebox{\zeroset}
\newsavebox{\fullline}

\begin{document}

\maketitle

\begin{abstract}
I discuss how to perform fault-tolerant quantum computation with
concatenated codes using local gates in small numbers of dimensions.
I show that a threshold result still exists in three, two, or one
dimensions when next-to-nearest-neighbor gates are available, and
present explicit constructions.  In two or three dimensions, I also
show how nearest-neighbor gates can give a threshold result.  In all
cases, I simply demonstrate that a threshold exists, and do not
attempt to optimize the error correction circuit or determine the
exact value of the threshold.  The additional overhead due to the
fault-tolerance in both space and time is polylogarithmic in the error
rate per logical gate.
\end{abstract}

\section{Introduction}

Quantum computation has the potential to offer vast speedups over
classical computation.  For instance, Shor's factoring
algorithm~\cite{Shor:factor} and Grover's database search
algorithm~\cite{Grover} both offer great improvements over classical
algorithms for the corresponding problems.  However, quantum computers
are likely to be highly susceptible to errors, whether caused by
imperfect gates, decoherence due to interactions with the environment,
or any other cause.

In classical computers, error correction is rarely necessary because
the classical bits are stored using digital devices, which, at every
time step, will restore the state of the system to a 0 or a 1.  They
also are made up of a large number of smaller particles (electrons,
usually), and therefore act as a simple (classical) repetition code.

The theory of quantum fault-tolerant
computation~\cite{Shor:FT,KL,KLZ,Kitaev,AB,Gottesman,Preskill,GP,Zalka}
has developed in an attempt to allow a similar remedy to the buildup
of errors in a quantum computer.  Instead of performing our algorithm
on some number of physical quantum bits ({\em qubits}), we implement
it on a collection of {\em logical} qubits encoded using a somewhat
larger number of physical qubits.  The logical qubits live in a
carefully chosen coding subspace of the full Hilbert space of the
physical qubits.  Then, by repeatedly performing an error correction
algorithm during the computation, we can at every step restore (or
nearly so) the data to the coding space, thus fending off errors.

In fact, by using concatenated quantum codes, we can produce an error
threshold~\cite{KL,KLZ,Kitaev,AB,GP}: if the fundamental error rates
per gate and per time step are below some threshold, we can perform
arbitrarily long quantum computations with arbitrarily low logical
error rate.  This threshold is known to be at least
$10^{-6}$~\cite{KLZ,AB}, but is probably at least a few orders of
magnitude higher~\cite{GP}.  Estimates range as high as
$1/700$~\cite{Zalka}.

However, this result relies on a number of assumptions about the
computer and errors, some of which are described below.  One
assumption is that gates can be performed between any pair of qubits.
In practice, this may not be practical at all, since in a large
computer, qubits will be constrained by the dimensionality of space to
be far apart from each other.  For instance, in Kane's proposal for a
solid-state quantum computer using single-spin NMR~\cite{Kane}, only
adjacent qubits (in one or perhaps two dimensions) can directly
communicate.

Luckily, this assumption is not critical to the result.  As Aharonov
and Ben-Or note~\cite{AB}, ``the procedures \ldots can be made \ldots
to operate only on nearest neighbors, by adding gates that swap
between qubits.''  As we shall see, this is true, but we must be
careful in how we design the computer.  The ancilla qubits we need to
perform error correction must be placed sufficiently near the
computational qubits to be corrected, or too many errors will
accumulate in the time necessary to move the interacting qubits
together, taking us above the error threshold.  Furthermore, when a
number of levels of concatenation are used (as is necessary for long
computations), some ancillas will necessarily be far away from the
data, and we must be certain this does not destroy the threshold
result.

\section{The Threshold Result}

First, I will review the usual threshold result.  Each qubit is
encoded with a concatenated quantum code, usually using the
seven-qubit code~\cite{Steane:7qubit}.  That is, each qubit is encoded
as seven qubits via the mapping
\begin{eqnarray}
\ket{0} & \mapsto & \ket{0000000} + \ket{1111000} + \ket{1100110} + 
\ket{1010101} + \nonumber \\
& & \ket{0011110} + \ket{0101101} + \ket{0110011} + \ket{1001011} \\
\ket{1} & \mapsto & \ket{1111111} + \ket{0000111} + \ket{0011001} + 
\ket{0101010} + \nonumber \\
& & \ket{1100001} + \ket{1010010} + \ket{1001100} + \ket{0110100},
\end{eqnarray}
and each of those seven qubits is again encoded using the same map,
and so on for $L$ levels.

The seven-qubit code has a number of properties that make it
particularly favorable for fault-tolerant computation.  The logical
$0$ of the seven-qubit code is the superposition of the even codewords
of the (classical) Hamming code, whereas the logical $1$ is the
superposition of the odd codewords of the Hamming code.  Therefore, to
make a measurement of $0$ vs.\ $1$, we need only measure each qubit in
the block individually, and we will be able to determine the
measurement result from the parity of the resulting Hamming
codeword~\cite{Preskill}.  We do not need to perform an additional
quantum error correction step before this measurement --- phase errors
will not affect the measurement result, and bit flip errors will show
up as bit flip errors in the classical codeword, which can be
corrected using classical methods.  (The parity of the codeword should
only be determined {\em after} this correction.)

In addition, it is easy to perform a number of fault-tolerant
operations on the seven-qubit code.  The Hadamard transform $H:
\ket{j} \mapsto (\ket{0} + (-1)^j \ket{1})/\sqrt{2}$, the phase gate
$P: \ket{j} \mapsto i^j \ket{j}$, and the controlled NOT (or XOR)
$\ket{j,k} \mapsto \ket{j, j \oplus k}$ can all be performed via
simple {\em transversal} operations.  A transversal operation only
involves gates that interact the $r$th qubit in a block with itself
and the $r$th qubits in other blocks.  This prevents any errors from
spreading within a block, so a single physical error cannot cause a
whole block of seven to go bad.

The logical CNOT has another useful property: individual (or multiple)
bit flip errors in the control block will propagate forwards,
producing the corresponding bit flip errors in the target block, and
phase errors in the target block will propagate backwards, producing
the corresponding phase errors in the control block.  In addition, the
logical Hadamard transform will convert bit flip errors to phase
errors and vice-versa, without changing the location of the errors.

\begin{figure}
\centering
\begin{picture}(100,80)

\put(0,14){\makebox(40,12){$\ket{0}$}}
\put(0,34){\makebox(40,12){$\ket{0} + \ket{1}$}}
\put(0,54){\makebox(40,12){data}}

\put(40,20){\line(1,0){34}}
\put(86,20){\line(1,0){14}}
\put(40,40){\line(1,0){60}}
\put(40,60){\line(1,0){60}}

\put(60,20){\circle*{4}}
\put(60,20){\line(0,1){44}}
\put(60,60){\circle{8}}

\put(80,60){\circle*{4}}
\put(80,60){\line(0,-1){24}}
\put(80,40){\circle{8}}

\put(74,14){\framebox(12,12){H}}

\end{picture}
\caption{Error correction circuit for the seven qubit code.  Each line
represents a block of seven qubits, and each gate represents the same
gate applied transversally on the block.}
\label{fig:ECcircuit}
\end{figure}
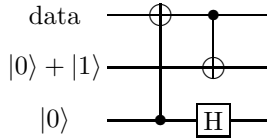

We can take advantage of these facts to produce a simple
fault-tolerant error correction circuit, shown in
figure~\ref{fig:ECcircuit}.
We use the CNOT to copy the bit flip errors from the data block into
an ancilla block, and measure the ancilla to see where those errors
are.  A similar procedure tells us the phase errors.  The ancillas
begin in the logical $\ket{0}$ and logical $\ket{0} + \ket{1}$ states
so that when the data is correct, it is unaffected by the error
correction procedure.  From just one measurement, there is no way to
tell whether an error is originally native to the data or to the
ancilla, so we should repeat this procedure multiple times.  Using
some decision process, we then decide what the most likely error is
and correct it.

Also, note that while we are measuring phase errors, bit flip errors
can pass from the ancilla into the data, and vice-versa.  If the
errors are just single-qubit errors, this is not a major problem ---
we can correct them using the regular error-correction procedure.
However, the process of creating the ancilla blocks may introduce
correlated errors, and if those errors enter the data, it will be a
serious problem.  Therefore, we must also verify the ancilla blocks to
eliminate such correlated errors.  Precisely how we do this is not
important for the discussion below, but it will certainly involve a
number of additional ancilla qubits.

It is not difficult to see that all of the above properties hold
equally well for the concatenated seven-qubit code.  CNOTs, Hadamard,
and phase gates can all be performed transversally on blocks of size
$7^L$, and error correction can be performed by interacting a full
block with additional ancilla blocks of size $7^L$.  Since the
concatenated seven qubit code is still a superposition of concatenated
classical Hamming codewords, we can determine the error at all levels
just from measuring these $7^L$-qubit blocks~\cite{Preskill}.

In order to complete a universal set of gates, we need an additional
gate, such as the Toffoli gate~\cite{Shor:FT} ($\ket{i, j, k} \mapsto
\ket{i, j, k \oplus ij}$).  The construction of the Toffoli gate is
somewhat involved, but requires three additional ancilla blocks
encoded at the same level as the data, plus some ``cat'' states
($\ket{00\ldots0} + \ket{11\ldots1}$) encoded with one less level.
The construction also uses a number of Toffoli gates performed at the
next lower level of concatenation.  Therefore, construction of the
Toffoli gate at level $L$ will first require the construction of
Toffoli gates at level $L-1$, which requires the construction of
Toffoli gates at level $L-2$, and so on.

Once we have all these tools, if we make a few assumptions about the
form of the errors and our capabilities, we can prove the existence of
an error threshold, below which arbitrarily long quantum computations
are possible.  At level $k$, there are a fixed number of places for
errors to occur, and two errors are required to produce an error at
level $k+1$ (our use of fault-tolerant procedures ensures this).
Therefore, the error rate $P_{k+1}$ at level $k+1$ is related to the
error rate $P_k$ at level $k$ by
\begin{equation}
P_{k+1} = C P_k^2,
\end{equation}
for some constant $C$, which essentially counts all possible
combinations of level $k$ errors that are fatal at level $k$ (though
most such combinations will be correctable at level $k+1$).  Then
\begin{equation}
P_L = (1/C) (C P_0)^{2^L}.
\end{equation}
$P_0$, the error rate at level $0$, is the error rate on the physical
qubits.  If $P_0 < 1/C$, then $P_L$ will approach $0$ extremely
rapidly, as a double exponential in the number of levels.  In fact,
since the number of qubits required to encode at level $L$ is
exponential in $L$, the error rate is a single exponential in the
number of qubits (or equivalently, the number of qubits required is
polylogarithmic in the desired error rate per gate).  The value of the
threshold is (at least) $1/C$.

Because this scaling is so rapid, the result still holds true even if
the number of possible places for errors increases, even exponentially,
with the number of levels.  Suppose
\begin{equation}
P_{k+1} = C r^{k+1} P_k^2,
\label{exponential}
\end{equation}
for constants $C$ and $r$.  Then
\begin{equation}
P_L = \frac{1}{C r^2} \left(C r^2 P_0 \right)^{2^L} / r^L.
\end{equation}
The threshold is still present, but is now $1/(C r^2)$.  This fact
will be key to showing a threshold is still present when we use local
gates.  In this case, we cannot avoid the error rate per level
increasing, but we will arrange the computer so that it only increases
according to equation~(\ref{exponential}).  Note that the final error
rate is still a double exponential in the number of levels.

This does not completely prove the existence of a threshold for
full-fledged fault-tolerant computation.  It also must be possible to
reliably create ancilla states for error correction at level $L$, and
it must be possible to perform Toffoli gates at level $L$.  Creating a
reliable level $L$ ancilla requires first a number of reliable level
$L-1$ ancillas, then a number of level $L-1$ gates.  For instance, to
create an encoded $\ket{0}$ state at level $L$, we take $7$ level
$L-1$ encoded $\ket{0}$ states and perform an appropriate circuit
interacting them to produce the level $L$ state.

Since multiple blocks (of $7^{L-1}$ qubits) within the level $L$ block
interact, this encoding network could introduce multiple correlated
errors in the level $L$ block, so we must verify the encoded states.
For instance, one way to do this is to compare them against each other
(using essentially the regular error-correction procedure) ---
correlated errors in one block will be uncorrelated with errors in
another block.  This will produce a smaller number of more reliable
ancillas, which we again compare against each other.  With a constant
number of rounds of comparison, we can reduce the chance of correlated
errors to less than the chance of a similar number of errors arising
individually (recall that {\em uncorrelated} errors are arising during
the verification procedure as usual).  Other methods will produce
qualitatively similar results.

If the effective error rate in the level $L-1$ ancillas and the level
$L-1$ gates is low enough, then we can create reliable level $L$
ancillas~\cite{GP}.  The procedure is self-similar --- if creating a
reliable ancilla at level $k$ (including all verification steps)
requires a total of $N_a$ level $k-1$ ancillas,\footnote{In the
example verification procedure above, $N_a$ would need to include not
just the seven level $k-1$ ancillas encoded for the original block,
but also all of the ancillas used to create the blocks against which
the original level $k$ ancilla is compared, and for the blocks against
which they are compared, and so on.} a level $k+1$ ancilla also
requires $N_a$ level $k$ ancillas.  Therefore, to create a level $L$
ancilla ultimately requires $N_a^L$ level $0$ ancillas.

Essentially the same logic applies to the level $L$ Toffoli gate.  The
requirements for reliable preparation and reliable Toffoli gates will
both somewhat lower the threshold, but will not destroy it.

Perhaps the most important assumption we make is that we can perform
operations in parallel.  Otherwise the situation becomes akin to
spinning plates to keep them from falling --- each requires a certain
amount of time to spin, and once the number of plates becomes too
large, we will not have time to spin all of them before the first one
falls.  The precise degree of parallelism we assume will heavily
impact the error threshold.  Generally I will assume maximum
parallelism --- all pairs of qubits that can interact can do so at the
same time, provided no qubit participates in two different
interactions at once --- but since I will not be calculating an
explicit error threshold, this assumption will not greatly impact the
discussion.

Another important assumption is that errors are uncorrelated, both in
time and in space.  If there is a chance $p$ per time step that the
whole computer will break down, we will not be able to perform more
than about $1/p$ computational steps on average.  We can allow
small-scale correlations without much damage (though the lower levels
of the code will be less effective than expected in that case), but
long-range correlations in the errors have the potential to cause
serious problems.  Systematic gate errors can also be tolerated, but
may significantly lower the threshold~\cite{KLZ}.

We also require a supply of fresh qubits during the computation to act
as ancillas during error correction.  These ancillas provide a place
for us to dump entropy --- otherwise the Second Law of Thermodynamics
would forbid arbitrarily long computations.  In this paper, I shall
assume that qubits can be initialized and erased in place.  This
appears to be a strict requirement --- if a qubit has to move a long
distance from where it is initialized to where it is used, it will
likely be randomized by the time it arrives.

Note that all three of the above assumptions apply equally well to
fault-tolerant {\em classical} computation.  The last could
conceivably be circumvented by a careful use of irreversible gates,
but the ability to perform an appropriate variety of gate is really
just another form of the same assumption.

Some other assumptions are useful, but not necessary.  For instance,
we generally assume that errors may randomize the qubits, but will
never cause them to leave the computational space.  Since it is
possible in principle to detect such a ``leakage,'' we can remove this
assumption by adding in ``stop leak'' gates that watch for such an
error.  It is also frequently convenient to assume we have the
capability to make measurements on the qubits during the computation,
and to rapidly perform (modest) classical computations between quantum
steps.  We can remove this assumption by simulating the classical
computation with a quantum circuit (though it must follow the design
principles of a classical fault-tolerant computer).  In the case of
local quantum gates, we would intersperse quantum bits with regions
designated for these classical computations (see, for instance,
\cite{Gacs} for a one-dimensional classical fault-tolerant
architecture).  Having a reliable classical computer available
considerably simplifies the task of building a fault-tolerant quantum
computer, since we can assume our decision processes (such as for
which error occurred) are reliable.

Another unnecessary assumption is that arbitrary pairs of qubits can
communicate {\em directly}.  In this paper, I shall show that it is
sufficient to interact nearby pairs of qubits.  Then by moving the
data around, we can allow originally distant pairs of qubits to
interact.  We must, however, be careful that the time required to do
so is not too large, or by the time the qubits are brought together,
they will both be erroneous.

\section{Swapping Qubits}

In order to perform effectively long-range interactions, we shall
require the ability to move qubits around.  We will accomplish this by
swapping adjacent qubits.  An appropriate series of swaps between
adjacent qubits will allow us to perform an arbitrary permutation of
the qubits.  We will primarily be interested in cyclic rotations,
moving a single qubit a distance $d$ (qubit $1$ becomes qubit $d$,
while qubit $s$ becomes qubit $s-1$ for $s = 2, \ldots, d$), requiring
$d-1$ swaps.  To interact two qubits initially a distance $d+1$ apart,
we perform one such rotation, bringing the first qubit adjacent to the
second, then interact the two, then perform a second rotation,
returning the first qubit to its original location.  Altogether, we
need $2(d-1)$ swaps for this interaction.

\begin{figure}
\centering
\begin{picture}(235, 90)

\put(20,14){\makebox(40,12){B}}
\put(20,39){\makebox(40,12){Aux.}}
\put(20,64){\makebox(40,12){A}}

\put(60,70){\line(1,0){10}}
\put(60,45){\line(1,0){10}}

\put(70,70){\line(1,-1){25}}
\put(70,45){\line(1,1){11}}
\put(95,70){\line(-1,-1){11}}

\put(60,20){\line(1,0){67}}
\put(95,70){\line(1,0){5}}

\put(100,70){\line(1,-1){23}}
\put(127,43){\line(1,-1){23}}
\put(126,20){\line(1,1){10}}
\put(140,34){\line(1,1){10}}
\put(153,47){\line(1,1){23}}

\put(95,45){\line(1,0){65}}
\put(150,20){\line(1,0){10}}
\put(176,70){\line(1,0){19}}

\put(160,45){\line(1,-1){25}}
\put(160,20){\line(1,1){11}}
\put(185,45){\line(-1,-1){11}}

\put(185,20){\line(1,0){10}}
\put(185,45){\line(1,0){10}}

\put(195,14){\makebox(40,12){A}}
\put(195,39){\makebox(40,12){Aux.}}
\put(195,64){\makebox(40,12){B}}

\end{picture}
\caption{Network to fault-tolerantly swap two computational qubits}
\label{fig:swap}
\end{figure}
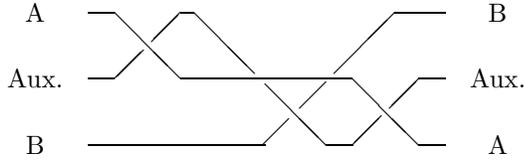

While we do not need to worry about the swap operation propagating
preexisting errors (it swaps the errors along with the data), we do
have to worry about errors in the swap gate itself, which could
introduce correlated errors in the two qubits being swapped.  To solve
this problem, we introduce an auxiliary qubit between the two
computational qubits A and B (which may themselves be ancillary to the
primary computation).  Then the following series of gates will swap A
and B without ever interacting them directly:
\begin{enumerate}
\item Swap (1, 2)
\item Swap (1, 3)
\item Swap (2, 3)
\end{enumerate}
(see figure~\ref{fig:swap}).
While the auxiliary qubit may acquire correlated errors with A or B,
that does not matter, since the value of the auxiliary qubit is
completely immaterial.\footnote{We might as well perform two unrelated
quantum computations on this computer at the same time, with the
auxiliary qubits for one being the computational qubits for the other.}
Note that this network requires next-to-nearest-neighbor gates.

\begin{figure}
\centering

\begin{picture}(300,60)

\savebox{\xline}(100,0){%
\multiput(10,0)(20,0){5}{\makebox(0,0){\textsf{x}}}
\multiput(20,0)(20,0){4}{\makebox(0,0){\textsf{o}}}}

\savebox{\oline}(100,0){%
\multiput(10,0)(20,0){5}{\makebox(0,0){\textsf{o}}}}

\put(0,45){\makebox(0,0){a)}}
\multiput(15,10)(0,20){3}{\usebox{\xline}}
\multiput(15,20)(0,20){2}{\usebox{\oline}}

\savebox{\oline}(100,0){%
\multiput(10,0)(20,0){5}{\makebox(0,0){\textsf{o}}}
\multiput(20,0)(20,0){4}{\makebox(0,0){\textsf{c}}}}

\put(185,45){\makebox(0,0){b)}}
\multiput(200,10)(0,20){3}{\usebox{\xline}}
\multiput(200,20)(0,20){2}{\usebox{\oline}}

\end{picture}

\begin{picture}(320, 130)


\put(0,85){\makebox(0,0){c)}}

\multiput(20,50)(0,40){2}{\makebox(0,0){\textsf{x}}}
\multiput(20,30)(0,40){2}{\makebox(0,0){\textsf{o}}}
\put(20,10){\makebox(0,0){\textsf{A}}}
\multiput(40,10)(0,40){3}{\makebox(0,0){\textsf{o}}}
\multiput(40,30)(0,40){2}{\makebox(0,0){\textsf{c}}}

\multiput(20,45)(0,40){2}{\vector(0,-1){10}}
\multiput(20,35)(0,40){2}{\vector(0,1){10}}
\multiput(25,30)(0,40){2}{\vector(1,0){10}}
\multiput(35,30)(0,40){2}{\vector(-1,0){10}}


\put(50,50){\vector(1,0){50}}

\multiput(110,50)(0,40){2}{\makebox(0,0){\textsf{o}}}
\multiput(110,30)(0,40){2}{\makebox(0,0){\textsf{c}}}
\put(110,10){\makebox(0,0){\textsf{A}}}
\multiput(130,10)(0,40){3}{\makebox(0,0){\textsf{o}}}
\multiput(130,30)(0,40){2}{\makebox(0,0){\textsf{x}}}

\multiput(110,25)(0,20){4}{\vector(0,-1){10}}
\multiput(110,15)(0,20){4}{\vector(0,1){10}}


\put(140,50){\vector(1,0){50}}

\multiput(200,10)(0,40){2}{\makebox(0,0){\textsf{c}}}
\multiput(200,30)(0,40){2}{\makebox(0,0){\textsf{o}}}
\put(200,90){\makebox(0,0){\textsf{A}}}
\multiput(220,10)(0,40){3}{\makebox(0,0){\textsf{o}}}
\multiput(220,30)(0,40){2}{\makebox(0,0){\textsf{x}}}

\multiput(200,25)(0,40){2}{\vector(0,-1){10}}
\multiput(200,15)(0,40){2}{\vector(0,1){10}}
\multiput(205,30)(0,40){2}{\vector(1,0){10}}
\multiput(215,30)(0,40){2}{\vector(-1,0){10}}


\put(230,50){\vector(1,0){50}}

\multiput(290,10)(0,40){2}{\makebox(0,0){\textsf{x}}}
\multiput(290,30)(0,40){2}{\makebox(0,0){\textsf{c}}}
\put(290,90){\makebox(0,0){\textsf{A}}}
\multiput(310,10)(0,40){3}{\makebox(0,0){\textsf{o}}}
\multiput(310,30)(0,40){2}{\makebox(0,0){\textsf{o}}}

\multiput(295,30)(0,40){2}{\vector(1,0){10}}
\multiput(305,30)(0,40){2}{\vector(-1,0){10}}

\end{picture}

\caption{a) Computational qubits (\textsf{x}) arranged on a square
lattice, interspersed by auxiliary qubits (\textsf{o}).  b) adds
cul-de-sacs (\textsf{c}) for moving qubits out of the way.  c) Moving
a data qubit (\textsf{A}) two positions.}
\label{fig:lattice}
\end{figure}
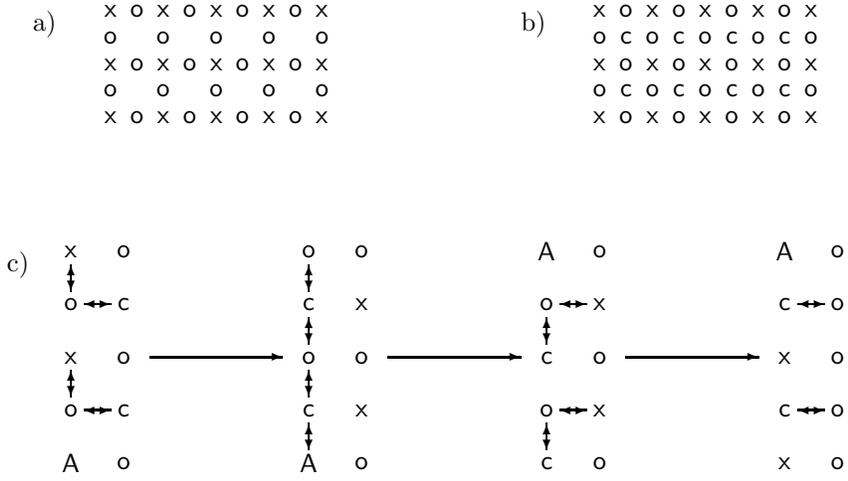

To allow swaps between arbitrary pairs of neighboring computational
qubits, we need to alternate computational qubits with auxiliary
qubits, as in figure~\ref{fig:lattice}a.  Note that in two or more
dimensions, we can manage with simple nearest-neighbor gates by
arranging cul-de-sacs where we can temporarily store one computational
qubit while moving another past it (figure~\ref{fig:lattice}b).  For
instance, to move a qubit A two positions up
(figure~\ref{fig:lattice}c), we simply slide the two (computational)
qubits above it into the cul-de-sacs down and to the right from their
normal positions, using regular nearest-neighbor swaps.  Then move A
up to its destination, and move the two displaced qubits into the
computational slots down and to the left.  With or without
cul-de-sacs, moving a qubit a distance $d$ requires $O(d)$ gates.

The remainder of the protocol only requires swaps and other
interactions between nearest neighbors.  One might think the Toffoli
gate would require next-to-nearest neighbor interactions, but in fact,
it can be built up from one- and two-qubit gates~\cite{gates}.
Therefore, in two or more dimensions, we will be able to perform
fault-tolerant computation with only nearest-neighbor interactions,
whereas in one dimension, the inability to use cul-de-sacs to move the
data out of the way requires us to go to next-to-nearest-neighbor
interactions.  However, in many {\em almost} one-dimensional systems
(such as two parallel lines of qubits, or a single line with an
occasional additional qubit on the side), we can again move to
nearest-neighbor interactions.

In fact, since swap gates cannot propagate errors, it will likely be
possible to use nearest-neighbor gates even in one-dimensional
systems.  However, since a single swap gate could introduce correlated
errors on pairs of qubits in the same block, it might be necessary to
use a concatenated code that corrects two errors per level instead of
the concatenated seven-qubit code.

\section{Three Dimensions}

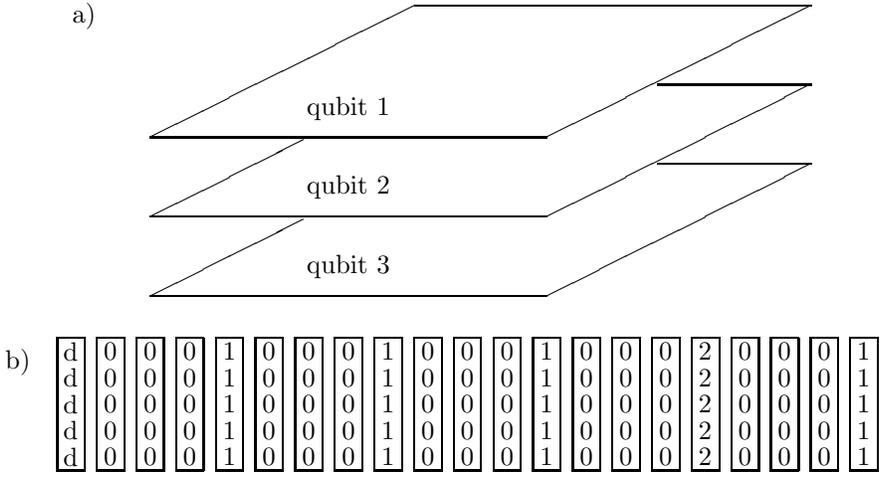
\begin{figure}

\centering

\begin{picture}(300,130)

\put(0,110){\makebox(30,12){a)}}

\multiput(40,70)(100,50){2}{\line(1,0){150}}
\multiput(40,70)(150,0){2}{\line(2,1){100}}

\put(115,75){\makebox(0,12){qubit 1}}

\put(40,40){\line(2,1){58}}
\put(40,40){\line(1,0){150}}
\put(190,40){\line(2,1){100}}
\put(290,90){\line(-1,0){58}}

\put(115,45){\makebox(0,12){qubit 2}}

\put(40,10){\line(2,1){58}}
\put(40,10){\line(1,0){150}}
\put(190,10){\line(2,1){100}}
\put(290,60){\line(-1,0){58}}

\put(115,15){\makebox(0,12){qubit 3}}

\end{picture}

\begin{picture}(330, 60)

\savebox{\zeroset}(30,0){%
\multiput(0,0)(15,0){3}{\makebox(0,0){0}}}

\savebox{\fullline}(300, 0){%
\put(0,0){\makebox(0,0){d}}
\multiput(60,0)(60,0){3}{\makebox(0,0){1}}
\put(240,0){\makebox(0,0){2}}
\put(300,0){\makebox(0,0){1}}
\multiput(15,0)(60,0){5}{\usebox{\zeroset}}}

\put(0,40){\makebox(10,12){b)}}
\multiput(25,10)(0,10){5}{\usebox{\fullline}}

\multiput(20,5)(15,0){21}{\framebox(10,50){}}

\end{picture}
\caption{a) The logical qubits of the computer lie on separate planes.
b) Each plane has the data on one line, adjacent to ancillas at
various levels.  The letter d represents a data qubit, 0 is a qubit
from a level 0 ancilla, 1 is from a level 1 ancilla, and so on.}
\label{fig:3D}
\end{figure}

When our qubits lie on a three-dimensional cubic lattice, we use the
arrangement of figure~\ref{fig:3D} for our computer.
Each logical qubit has associated with it many ancillas.  We arrange a
single encoded data qubit on a plane with all of its ancillas.  Then
we stack the planes, aligning the data qubits in all of the planes.
Therefore, to perform a tranversal interaction between adjacent data
qubits, we do not need to move anything.  To perform a transversal
interaction between distant data qubits, we must first move them
together.  By treating each step in the move as a regular
computational step, we can see that a computation involving $K$
logical qubits will be slowed down by a factor of at most $K$
computational steps.  When $K$ is large, we will have to perform error
correction at intermediate stages during the move, but this does not
present any particular extra burden, since we have the ancillas for
error correction constantly available.

In a single plane, we have many lines of qubits.  One line will
consist of the data block of $7^L$ qubits.  Other lines will consist
of ancillas of various forms and functions.  After being used, an
ancillas is reinitialized to be used again (a process which we are
assuming can be done in place).  The ancillas are aligned with the
data so that all interactions take place along a single
``interaction'' axis (perpendicular to the arrangement of the data
block), thus realizing the transversal nature of the interactions.
The only time interactions will occur along the other axis (the
``data'' axis) is when an ancilla is being encoded (to create a
``cat'' state, for instance, or a level $k$ ancilla from level $k-1$
ancillas).  Any such encoding will always be followed by verification
steps.

The ancillas necessary for level $0$ operations, including both error
correction and the Toffoli gate, must be nearby the data --- we cannot
tolerate moving qubits a long distance for level $0$ operations.
Therefore we place the level $0$ ancillas immediately adjacent to the
data along the interaction axis.  There are a fixed number $N_o$ of
such ancillas, not increasing with the number of levels of encoding
used in the computer (we only count the ancillas that {\em directly}
interact with the data qubit).  Depending on their function, these
level $0$ ancillas may have a number of different forms.  They are
frequently encoded in blocks of $7$, which interact with corresponding
$7$-qubit blocks of the code.  Level $0$ ancillas which are adjacent
along the data axis are completely independent --- they interact with
different blocks of the data, and no communication between them is
necessary.

There are also a total of $N_o$ level $1$ ancillas for level $1$
operations and error correction on the data.  We place the first of
these after the last level $0$ ancilla on the interaction
axis.\footnote{Note that a ``level $k$ ancilla'' is used for
operations or error correction at level $k$.  When it is used for
error correction, it may in fact be a state encoded at level $k+1$.}
Following the first level $1$ ancilla, we place a number $N_t$ of
level $0$ ancillas, which are necessary for preparation {\em and
error-correction} of the level $1$ ancilla.  After that comes another
level $1$ ancilla, followed by another set of level $0$ ancillas, and
so on.  We require the ability to correct errors on the level $1$
ancillas because they may have to move a considerable distance to
interact with the data, and we may wish to preserve their state during
the move.

After all of the level $1$ ancillas, we place the first level $2$
ancilla, followed by the level $0$ ancillas necessary to maintain it.
After those comes the first of the level $1$ ancillas necessary to
prepare and maintain the level $2$ ancilla, then the level $0$
ancillas to prepare and maintain the level $1$ ancilla, then another
level $1$ ancilla, and so on.  The level $2$ ancilla requires a total
of $N_t$ level 1 ancillas, and each has associated with it $N_t$ level
$0$ ancillas.  Therefore, each level $2$ ancilla requires at most
$(N_t+1)^2$ lines of qubits for its support structure.

We follow this pattern as far as necessary --- each level $k$ ancilla
requires $N_t$ level $k-1$ ancillas, each of which requires $N_t$
level $k-2$ ancillas, each of which requires $N_t$ level $k-3$
ancillas, and so on.  We can see that the total number of ancillas
grows exponentially with level.  This means that to interact with the
data qubit, a level $k$ ancilla will have to move a distance $N^k$ for
some constant $N$ (greater than $N_t$ and $N_o$).  This means that the
interaction takes $O(N^k)$ times as much time to occur as when we had
long distance interactions, so the possibilities for error also
increase exponentially with $k$.  However, this merely recovers
equation~(\ref{exponential}),
\begin{equation}
P_{k+1} = C r^{k+1} P_k^2,
\end{equation}
which still has an error threshold.  Therefore, we can perform
fault-tolerant computation with local gates in three dimensions.

Note that a block encoded at level $k$ may have different error rates
at level $0$, level $1$, and so on (a level $0$ error is a single
erroneous qubit; in a level $1$ error, a whole block of $7$ has gone
bad).  It is important that only the probability of level $k$ errors,
and not the probability of level $0$ errors (or errors at another
fixed level), increases with $k$.  If a level $k$ ancilla experiences
an exponential (or even linear) increase in the probability of level
$0$ errors, equation~(\ref{exponential}) will not be valid.  It is to
solve this potential problem that we need to be able to correct errors
on the ancillas as well as prepare them.  For instance, we might move
the level $k$ ancilla $s$ spaces towards the data, then perform error
correction using the local extra ancillas, staving off low-level
errors, then resume its movement.  Note that to perform level $l$
error correction on the ancilla (with $l < k$), we may need to move
level $l$ ancillas around to bring them next to the level $k$ ancilla.
However, these level $l$ ancillas will only have to move a distance at
most $N^l$,\footnote{Or perhaps $2 N^l$, since partway through the
move two level $k$ ancillas could be adjacent.} so
equation~(\ref{exponential}) holds.  We should also perform low-level
error correction on the data at the same time as on the ancilla, since
the data is accumulating errors while it waits for the level $k$
ancilla to arrive.  Some other ancillas may also contain important
information, and we should perform error correction on them too.

Since level $L$ ancillas now have to move a distance $N^L$ to interact
with the data, this protocol produces a slowdown by a factor of
$O(N^L)$ relative to the usual fault-tolerant protocol.  Since the
procedure requires more stops for low-level error correction, which in
turn means more ancillas, there will be a similar increase in the
number of qubits needed.  However, since the error rate is a double
exponential in $L$, this means the additional overhead is only
polylogarithmic in the desired error rate.

\section{Two Dimensions}

In two dimensions, we adopt a somewhat similar arrangement.  Now the
individual data qubits and their ancillas form lines, which are
again aligned so that transversal gates between adjacent data qubits
are straightforward (see figure~\ref{fig:2D}).
\begin{figure}
\centering

\begin{picture}(310,65)

\savebox{\zeroset}(20,0){%
\multiput(0,0)(10,0){3}{\makebox(0,0){0}}}

\savebox{\fullline}(270,0){%
\multiput(0,0)(40,0){3}{\makebox(0,0){1}}
\multiput(10,0)(40,0){3}{\usebox{\zeroset}}
\multiput(120,0)(40,0){3}{\makebox(0,0){d}}
\multiput(130,0)(40,0){3}{\usebox{\zeroset}}
\put(240,0){\makebox(0,0){1}}
\put(250,0){\usebox{\zeroset}}}

\multiput(50,10)(0,15){4}{\usebox{\fullline}}

\put(0,55){\makebox(40,0){qubit 1}}
\put(0,40){\makebox(40,0){qubit 2}}
\put(0,25){\makebox(40,0){qubit 3}}
\put(0,10){\makebox(40,0){qubit 4}}

\end{picture}

\caption{The logical qubits of the computer lie on separate lines.
Within each line, ancillas are interspersed with the data qubits.
Again, d represents a data qubit, 0 is from a level 0 ancilla, and 1
is from a level 1 ancilla.}
\label{fig:2D}
\end{figure}

The arrangement of the individual lines can also be seen in
figure~\ref{fig:2D}.  Next to each data qubit, we place the corresponding
qubits from the $N_o$ level $0$ ancillas.  We do this for a block of
seven data qubits (since we are using the seven-qubit code), and then
place the level $1$ blocks for the $N_o$ level $1$ ancillas required
to perform error correction and Toffoli gates at level $1$ with the
data.  These blocks themselves contain, interspersed with the qubits
in the level $1$ blocks, the $N_t$ level $0$ ancillas to create and
maintain the level $1$ ancillas.  We repeat this pattern (level $1$
data blocks next to $N_t$ level $1$ ancilla blocks) seven times, then
position the level $2$ ancilla blocks with their support structures
between the level $2$ data blocks.

As in the three dimensional case, this structure means that a level
$k$ ancilla will have to move a distance $N^k$ to interact with the
data (although $N$ may be bigger).  Again, we may have to perform
level $l$ error correction along the way (with $l < k$), but a level
$l$ ancilla is never further than $N^l$ places away.  We once again
arrive with a recursion relation in the form~(\ref{exponential}), so
we still have an error threshold.

\section{One Dimension}

Suppose we had just a two-qubit quantum computer.  We could easily
convert the two-dimensional model into a one-dimensional model by
alternating qubits from the line associated with the first data qubit
with the line associated with the second data qubit (so we would have
a qubit from data block $1$ followed by a qubit from data block $2$,
then a level $0$ ancilla qubit for data block $1$, then a level $0$
ancilla qubit for data block $2$, and so on).  In this model, each
ancilla will have to move exactly twice as far as in the
two-dimensional case, so there will still be a threshold, though it
will be half as large.  To interact the two data qubits, we should
perfectly align the blocks, so that qubit number $57$ from block $1$
is right next to qubit number $57$ from block $2$.  However, if the
logical qubits do not need to interact, there is no reason the blocks
need to be aligned.  We will still be able to perform error correction
on the blocks separately, even if they are out of phase.

In the two-dimensional model, each data block with its ancillas (and
the support structure for the ancillas) took up only a finite amount
of space ($T^L$ for some constant $T$ when there are $L$ levels
altogether).  That means that we can create an arbitrarily large
one-dimensional quantum computer by placing these blocks of $T^L$
alongside each other.

However, to interact two adjacent blocks would require moving qubits a
distance $T^L$ --- too far to go without error correction.  The
solution is to move the support structure for the data block along
with the data block itself, interleaving the two blocks as in the
two-qubit example.  We will probably have to stop during the move to
do error correction, and the blocks will still be out of phase at this
point.  Since we do not need to interact the blocks to perform error
correction, this is not a problem.  We can bring the blocks into
phase, interact them, then move them back apart.

All in all, this process will slow the computer down by an additional
factor of $T^L$ (beyond the two-dimensional case).  Again, this only
results in an additional polylogarithmic slowdown.  Therefore, even
for the one-dimensional case, fault-tolerant quantum computation is
possible with local gates.

\section*{Acknowledgements}

I would like to thank David DiVincenzo and particularly John Preskill
for helpful conversations.  This research is supported by the
Department of Energy under contract W-7405-ENG-36.


\begin{thebibliography}{99}
\bibitem{Shor:factor} P.~W.~Shor, ``Polynomial-time algorithms for
prime factorization and discrete logarithms on a quantum computer,''
{\it Proc. 35th Annual Symp. on the Foundations of Computer Science},
p. 124 (IEEE Computer Society Press, Los Alamitos, CA, 1994),
quant-ph/9508027.
\bibitem{Grover} L.~Grover, ``Quantum mechanics helps in searching for
a needle in a haystack,'' Phys. Rev. Lett. {\bf 79}, 325 (1997).
\bibitem{Shor:FT} P.~W.~Shor, ``Fault-tolerant quantum computation,''
{\it Proc. 37th Ann. Symp. on the Foundations of Computer Science},
p. 56 (IEEE Computer Society Press, Los Alamitos, CA, 1996),
quant-ph/9605011.
\bibitem{KL} E.~Knill and R.~Laflamme, ``Concatenated quantum codes,''
quant-ph/9608012.
\bibitem{KLZ} E.~Knill, R.~Laflamme, and W.~H.~Zurek, ``Threshold
accuracy for quantum computation,'' quant-ph/9610011; E.~Knill,
R.~Laflamme, and W.~H.~Zurek, ``Resilient quantum computation,''
Science {\bf 279}, 342 (1998); E.~Knill, R.~Laflamme, and W.~H.~Zurek,
``Resilient quantum computation: error models and thresholds,''
Proc. Royal Soc. Lond. A {\bf 454}, 365 (1998), quant-ph/9611025.
\bibitem{Kitaev} A.~Y.~Kitaev, ``Quantum error correction with
imperfect gates,'' {\it Quantum Communcation, Computing, and
Measurement (Proc. 3rd Int. Conf. of Quantum Communication and
Measurement)}, p. 181 (Plenum Press, New York, 1997).
\bibitem{AB} D.~Aharonov and M.~Ben-Or, ``Fault-tolerant quantum
computation with constant error,'' {\it Proc.\ 29th Ann.\ ACM Symp.\ on 
Theory of Computing}, p. 176 (ACM, New York, 1998), quant-ph/9611025.
\bibitem{Gottesman} D.~Gottesman, ``Theory of fault-tolerant quantum
computation,'' Phys.\ Rev.\ A {\bf 57}, 127 (1998), quant-ph/9702029.
\bibitem{Preskill} J.~Preskill, ``Fault-tolerant quantum
computation,'' ch. 8 in {\it Introduction to Quantum Computation and
Information}, eds. Hoi-Kwong Lo, Sandu Popescu, and Tim Spiller,
p. 213 (World Scientific, New Jersey, 1998), quant-ph/9712048.
\bibitem{GP} D.~Gottesman and J.~Preskill, unpublished.
\bibitem{Zalka} C.~Zalka, ``Threshold estimate for fault-tolerant
quantum computing,'' quant-ph/9612028.
\bibitem{Kane} B.~E.~Kane, ``A silicon-based nuclear spin quantum
computer,'' Nature {\bf 393}, 133 (1998).
\bibitem{Steane:7qubit} A.~M.~Steane, ``Error correcting codes in
quantum theory,'' Phys. Rev. Lett. {\bf 77}, 793 (1996).
\bibitem{Gacs} P.~G\'{a}cs, ``Reliable computation with cellular
automata,'' J. Comp. Sys. Sci. {\bf 32}, 15 (1986).
\bibitem{gates} A.~Barenco, C.~H.~Bennett, R.~Cleve, D.~P.~DiVincenzo,
N.~Margolus, P.~Shor, T.~Sleator, J.~A.~Smolin, and H.~Weinfurter,
``Elementary gates for quantum computation,'' Phys. Rev. A {\bf 52},
3457 (1995), quant-ph/9503016.
\end{thebibliography}
\end{document}